\documentclass[%
 reprint,
superscriptaddress,
 amsmath,amssymb,
 aps,
]{revtex4-2}

\usepackage{graphicx}
\usepackage{dcolumn}
\usepackage{bm}

\begin{document}

\preprint{APS/123-QED}

\title{Photonic Flatband Resonances in Multiple Light Scattering}

\author{Thanh Xuan Hoang}
 \email{hoangtx@ihpc.a-star.edu.sg}
\affiliation{Department of Electronics and Photonics, Institute of High Performance Computing, A$^\star$STAR (Agency for Science, Technology and Research), 1 Fusionopolis Way, \#16-16 Connexis, Singapore 138632, Singapore}%

\author{Daniel Leykam}%
\email{daniel.leykam@gmail.com}
\affiliation{Centre for Quantum Technologies, National University of Singapore, 3 Science Drive 2, Singapore 117543}%

\author{Yuri Kivshar}%
\email{yuri.kivshar@anu.edu.au}
\affiliation{Nonlinear Physics Center, Australian National University, Canberra ACT 2601, Australia}%

\date{\today}

\begin{abstract}
We introduce the concept of photonic flatband resonances for the example of an array of high-index dielectric particles. We employ  the multiple Mie scattering theory and reveal that both short- and long-range interactions between the resonators are crucial for the emerging {\it collective resonances} and their associated {\it photonic flatbands}. By examining both near- and far-field characteristics, we uncover how the flatbands emerge due to fine tuning of resonators' radiation fields, and predict that hybridization of a flatband resonance with an electric hotspot can lead to giant values of the Purcell factor for the electric dipolar emitters. 

\end{abstract}

\maketitle

{\it Introduction.---}  The physics of the systems with flat-band spectra (or flatbands) has attracted a lot of attention recently, in connection with the studies of electronic periodic structures with nontrivial topology~\cite{Sutherland1986,Leykam2018APX}. One of the important features of the flatband systems is that their density of states grows with the system size, in contrast to regular lattices where the density of states generally remains finite.  The enhanced density of states in flatbands allows achieving strong interaction in electronic systems ~\cite{doi:10.1142/S021797921330017X,Balents2020} and photonics~\cite{Leykam2018APL, Tang2020} with applications to quantum networks \cite{Reiserer2015,Higg2022}, on-chip single-photon generation \cite{Liu2018} and nanolasers \cite{Hoang2020}, as well as compact free-electron light sources \cite{Yang2023}. 

Importantly, a majority of the previous studies of flatbands involved the systems with engineered symmetries and couplings that are short-ranged either in real space~\cite{doi:10.1080/23746149.2021.1878057} or Fourier space~\cite{PhysRevLett.120.066102}, being described by the tight-binding models or coupled-mode theories. However, recent demonstrations of electronic and photonic moir\'{e} superlattices revealed that the flatbands may emerge from parameter fine-tuning in more general settings involving interactions between many states in the lattices with complex unit cells~\cite{PhysRevLett.99.256802,doi:10.1073/pnas.1108174108,Wang2020,Tang2021,PhysRevLett.126.223601,PhysRevLett.126.136101,PhysRevLett.128.253901,PhysRevResearch.4.L032031}. While the physics of flatbands remains an open question beyond the short-range coupling approximation, minimal effective tight binding models developed for magic angle bilayer graphene suggest that fine-tuned coupling between strongly- and weakly-localized states may play an important role~\cite{PhysRevB.99.195455,PhysRevResearch.1.033072}, in contrast to short-range interaction that typically destroys the flatbands.

We notice that couplings between states of very different nature occur routinely in photonics; while electronic systems described by the Schr\"odinger equation support bound states, light waves always have a positive energy and thus couple to an infinite number of radiation channels supported by the surrounding environment. Multipole-based models offer a powerful approach for describing couplings between localized and radiation states~\cite{Huang2011,doi:10.1098/rsta.2016.0317,PhysRevB.105.L241301}, but existing multipole models neglect effects such as the anisotropy of near-field scattering, coupling and interference between different multipole orders, and the infinite number of radiation channels in the surrounding three-dimensional (3D) environment, in part because they are expected to destroy the fine-tuned interference producing flatbands in simple effective models. 

\begin{figure}
\includegraphics[width = \columnwidth]{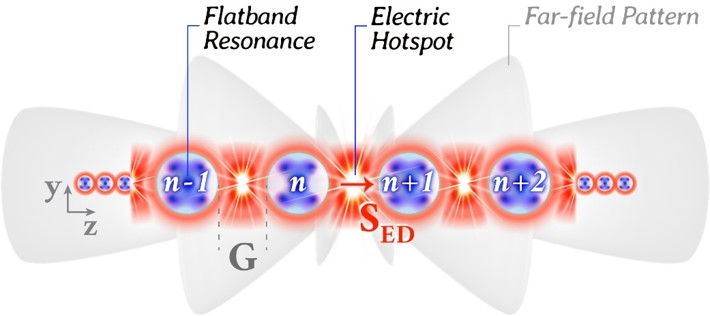}
\caption{\label{F1} Schematic of multipole Mie scattering for an electric dipole $S_{ED}$ interacting with a chain of dialectic spheres. The flatband resonance and electric hotspot modes are inseparable from the collective resonances and their far-field patterns.}
\end{figure}

In this Letter, we employ the multipole Mie scattering to demonstrate the emergence of photonic flatbands due to fine-tuning of interaction in a simple one-dimensional (1D) chain of high-index dielectric nanoparticles. Multiple scattering theory (MST) plays a crucial role in many fundamental wave phenomena, such as the Anderson localization of electrons and light~\cite{Lagendijk1996}. The MST approach allows to identify flatbands emerging as exact solutions to Maxwell's equations and study how their hybridization with electronic hotspot modes can lead to giant enhancement of the Purcell factor for electronic dipolar emitters~\cite{Hoang2017}. We find that coupling between different multipoles, which is typically considered in the context of short-range interactions~\cite{PhysRevLett.120.097401,PhysRevX.9.031010,PhysRevLett.128.256602}, also has a significant role in shaping the radiative losses~\cite{Hoang2017,Liu2020}. Most photonic flatbands have been based on fine-tuned couplings between different sublattices of resonators designed using tight binding models~\cite{Leykam2018APL}; our MST shows how flatband physics can emerge in much broader classes of resonator lattices due to competition between near- and far-field coupling.


{\it Model.---} We consider the interaction between a $z$-oriented electric dipolar emitter $S_{ED}$ with dipole moment $\mu$ and a linear chain of identical silicon spheres with refractive index 3.5, radius 210 nm, and separation $G$ embedded in vacuum, shown in Fig.~\ref{F1}. 
These parameters ensure the photonic modes of interest fall within the near-infrared frequency range, which is relevant for various scientific and technological applications \cite{Hoang2022} (see Supplementary Section S1 for relevant Mie modes and their associated electric hotspot modes). 
We expand the field scattered by the \textit{$u$}-th sphere positioned at ${\bf r}_u$ in terms of a series electric multipole fields ${\bf N}_{l;m}$~\cite{Hoang2017}
\begin{equation}
{\bf E}_u({\bf r}) = \sum_{l=1}^{L_u}p_{l;0}^{(u)}{\bf N}_{l;0}(k[{\bf r} - {\bf r}_u]), \label{eq1}
\end{equation}
where the required truncation order $L_u$ depends on the vacuum wavenumber $k$ and the spheres' radius; here $L_u = 10$ is sufficient to  obtain good agreement with a direct numerical solution of Maxwell's equations (Lumerical FDTD)~\cite{SuppMat}. Note that due to the axial symmetry the source excites only the $m=0$ electric multipole modes.

In the multiple Mie scattering theory the interaction between the dipolar emitter and the sphere, as well as the coupling between the sphere's scattering multipole expansion coefficient (MECs) is described by
\begin{equation}
p_{l^\prime;0}^{(n)}=a_{l^\prime}^{(n)}\left(A_{l^\prime;0}^{1;0}(\overrightarrow{OO_n})\mu+\sum_{u\neq n}\sum_{l=1}^{L_u}A_{l^\prime;0}^{l;0}(\overrightarrow{O_uO_n})p_{l;0}^{(u)}\right),\label{eq2}
\end{equation}            
where $A_{l^\prime;0}^{l;0}(\overrightarrow{O_uO_n})$ translates the multipole field of ${\bf N}_{l;0}$ from the $u$-th sphere into the incident field approaching the $n$-th sphere~\cite{Hoang2017}. Equation~\eqref{eq2} accounts for both the short-range ($u=n \pm 1$) and long-range couplings ($u\neq n\pm 1$) between the multiple scattering fields; it also includes the off-diagonal coupling between different multipolar orders, \textit{i.e.} $l^\prime \neq l$. Implicitly, Eq.~\eqref{eq2} also accounts for the coupling between the near field and the fields of an infinite number of plane waves supported by the 3D vacuum, since each multipole term ${\bf N}_{l;0}$ can be expanded into an infinite number of plane-wave modes~\cite{Hoang2014}.


After solving Eq.~\eqref{eq2} to obtain the individual spheres' MECs $p_{l;0}^{(u)}$, the far field can be obtained by summing the source field and the scattering fields from the spheres and performing a multipolar expansion of order $L_{\mathrm{tot}}$ about the origin using the vector addition theorem \cite{Chew1993,Hoang2017},
\begin{equation}
{\bf E}^{\mathrm{tot}}({\bf r}) = \sum_{l=1}^{L_{\mathrm{tot}}}\alpha_{l;0}{\bf N}_{l;0}(k{\bf r}), \label{eq3}
\end{equation}       
where the eigenfunctions ${\bf N}_{l;0}$ represent modes of the universe including the emitter, spheres, and their environment. The coefficients $\alpha_{l;0}$ implicitly take into account interference between the dipole source and scattered fields. The enhancement of the local density of states (LDOS) at $S_{ED}$ provided by the modes of Eq.~\eqref{eq3} is described by the Purcell factor
\begin{equation}
    F_P = \frac{1}{2|\mu|^2} \sum_{l=1}^{L_{\mathrm{tot}}} l(l+1) |\alpha_{l;0}|^2, \label{eq4}
\end{equation}
which is given by the total power radiating into the far field $P$ divided by the radiating power of an isolated $S_{ED}$ source, $P_0 = c|\mu|^2 / (4\pi)$. Thus, after obtaining the field generated by a dipolar source as a function of $k$ by solving Eq.~\eqref{eq2}, $F_P(k)$ can be computed using Eq.~\eqref{eq4} to identify resonances associated with excitations of the collective modes of the chain of nanospheres. 



\begin{figure}
\includegraphics[width = \columnwidth]{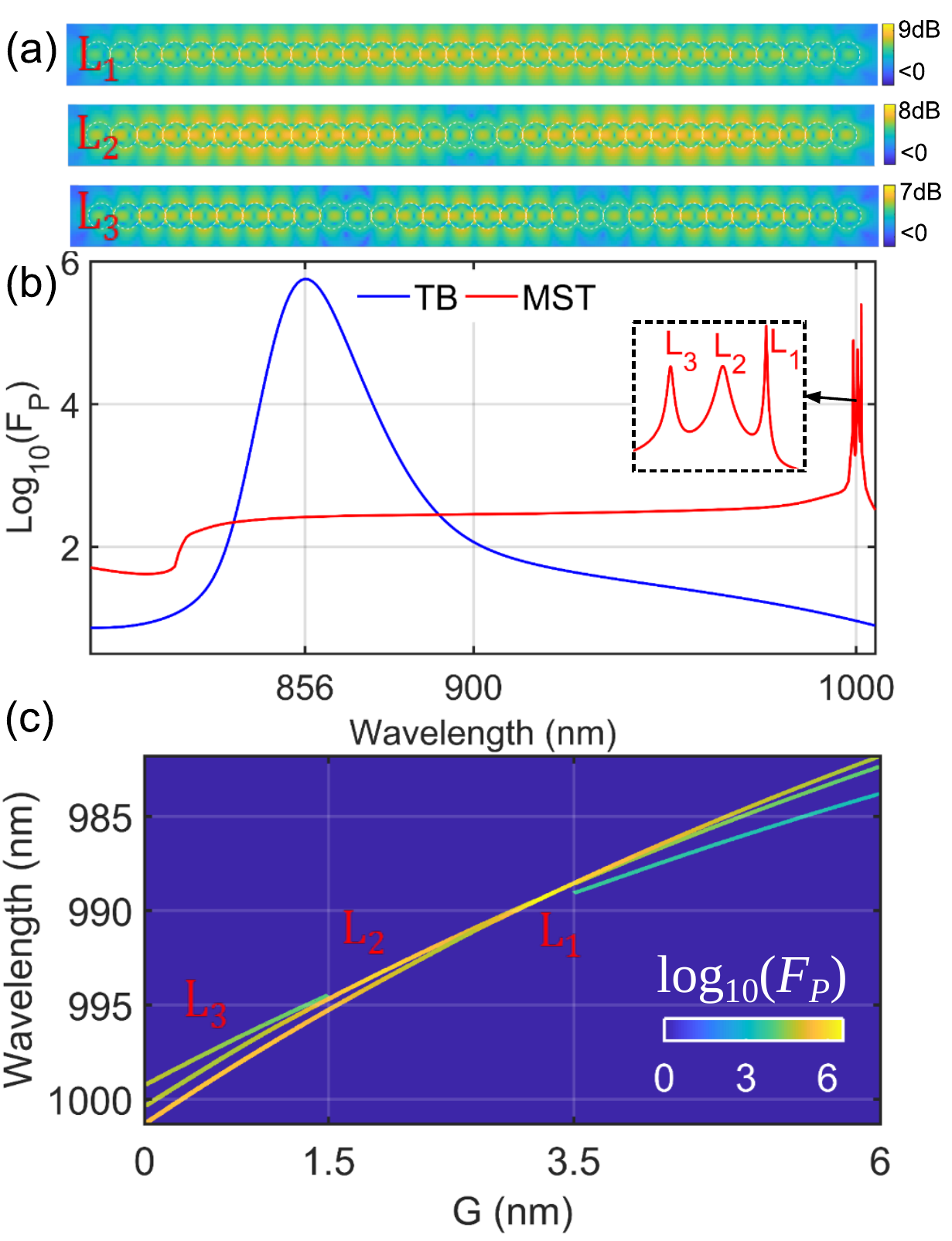}
\caption{\label{F2} Collective Mie resonances and their crossing regime. (a) Electric near-field distributions of the three collective quadrupole resonances excited by a dipolar emitter in a 30-sphere chain with the inter-particle gap size $G = 0$ nm. (b) The tight-binding (TB) model, which considers short-range couplings only, fails to capture the collective resonances present in the red line of the exact multiple scattering theory (MST). (c) Crossing and disappearance of the three collective $L_{1;2;3}$ resonances from the fine-tuning of $G$.}
\end{figure}

{\it Collective resonances.---} Figures \ref{F2}(a) and \ref{F2}(b) present the near-field distributions and $F_P$ spectrum of three collective $L_{1,2,3}$ resonances supported by a 30-sphere chain and excited by an $S_{ED}$ source located at the centre of the gap between the $7$-th and $8$-th spheres (see Supplementary Section S2 for the formation of the collective resonances).
Although the near-field distributions may appear similar to Fabry-Perot cavity modes \cite{Sidorenko2021,Kornovan2021}, it is important to note that the latter are due to propagating waves and only experience radiative losses at the ends of their cavity, as observed in nanowires \cite{Hoang2020}. In contrast, the radiative losses of the three $L_{1,2,3}$ resonances, described by Eq. \eqref{eq2}, occur along the entire chain, with the main contributors being the scatterings by the middle spheres (see Supplementary Section S3 for further discussion).

Figure \ref{F2}(b) demonstrates the essential role of long-range couplings in the physics of collective resonances. The blue line, which represents a tight binding-like model that neglects the long-range couplings in Eq.~\eqref{eq2} and considers only interactions between neighboring spheres, fails to capture the collective resonances observed in the exact MST model. These resonances are marked by the red peaks around the wavelength of $1000$ nm. Figure~\ref{F2}(b) also shows the significant impact of the electric hotspot mode on the LDOS, visible as a enhancement of $F_P$ by more than two orders of magnitude even for off-resonant wavelengths. Supplementary Section S4 presents the crucial role of the off-diagonal mode couplings in the MST, illustrating the inability to reproduce the collective resonances using a simpler model of interacting dipoles. The MST-based theory, equivalent to the electromagnetic-field Green's tensor \cite{Lodahl2015}, highlights the critical role played by both short- and long-range couplings in nanophotonics. Furthermore, the off-diagonal couplings between different multipole orders, inherent to the electromagnetic field Green's tensor, are also crucial to accurately describe the LDOS enhancement~\cite{Sauvan2013,Kristensen2014,Franke2019}.

Figure \ref{F2}(c) shows the peak $F_P$ values and corresponding wavelengths of the three collective resonances when fine-tuning $G$. As we increase $G$ from $0$ to $1.5$~nm, the weak $L_3$ resonance merges with and enhances the $L_2$ resonance to become even stronger than the $L_1$ resonance. The enhancement of the $L_2$ resonance comes at the cost of a complete suppression of the $L_3$ resonance for $G\in(1.5,3.5)$ nm. Increasing $G$ further, the $L_2$ resonance peaks at $G =1.8$ nm, where its $F_P$ is sixfold higher than that of the $L_1$ resonance (see Supplementary Section S5 for details). This $L_2$ resonance is also known as a super-cavity mode, since its linewidth depends strongly on the fine-tuning parameter $G$ near its optimal value \cite{Hoang2022}.

Feshbach's theory of overlapping resonances, which is described by a coupled-channel Schr\"{o}dinger equation, can explain the behavior of the collective resonances. When two closed channels couple to one open channel, a Friedrich-Wintgen bound state in the continuum (BIC) supported by a finite potential structure can be observed for anti-crossing modes~\cite{Friedrich1985,Hsu2016,Rybin2017}. The super-cavity $L_2$ resonance can be considered similar to a Friedrich-Wintgen BIC because its linewidth is much narrower at $G =1.8$ nm ($\Delta\lambda \approx 0.012$ nm, or equivalently a quality factor of $Q\approx 8.3\times 10^4$) than at $G = 0$ nm ($\Delta\lambda \approx 0.22$ nm, or $Q\approx 4.5\times 10^3$). We also observe a super-cavity $L_1$ resonance peaking at $G = 3.225$ nm with $Q \approx 1.4\times10^6$, where the $L_2$ resonance crosses into the $L_1$ resonance.

In contrast to BICs for matter waves, which in theory have a vanishing linewidth, light waves always have a positive energy and thus can couple to an infinite number of radiation plane-wave channels supported by the surrounding 3D vacuum, leading to a non-zero linewidth ($\Delta\lambda \neq 0$ nm). The Friedrich-Wintgen BIC theory assumes that two closed channels couple to one open channel, but this assumption is not valid for photonic systems due to the presence of the infinite number of radiation channels, approximated in terms of a finite number of multipole channels in multipole scattering theory~\cite{Hoang2014}. Feshbach's theory shows that one can always reduce these open channels to one continuum; eliminating the open channels results in an effective complex potential with an imaginary part that accounts for the energy loss associated with the eliminated open channels \cite{Friedrich1985}. 

The physics of Friedrich-Wintgen BICs for cases with complex potentials arising from many radiation channels remains unexplored, but it is analogous to the physics of the succession of crossing and anti-crossing resonances in Fig.~\ref{F2}(c) due to the analogy between matter and light waves with positive energies \cite{Lagendijk1996}. In particular, a complex potential can result in overlapping resonances with differing linewidths, corresponding here to the crossing at $G=1.5$ nm of the broader $L_3$ resonance which increases the quality factor of the $L_2$ resonance.

\begin{figure}
\includegraphics[width = \columnwidth]{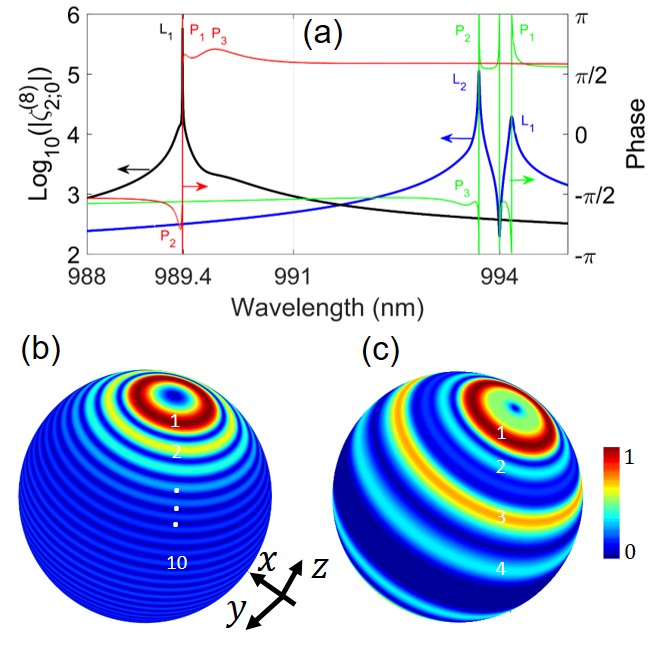}
\caption{\label{F3} (a) Magnitude and phase of the EQ coefficient $\zeta_{2;0}^{(8)}$ for the two cases of $G = 1.8$ nm (blue and green lines) and $G = 3.225$ nm (black and red lines). (b)-(c) Far-field patterns of the $L_1$ resonance for the two cases of $G = 0$ nm and $G = 3.225$ nm, respectively.}
\end{figure}

To provide a more comprehensive understanding of the collective resonances, we present their internal field and far-field characteristics in Fig. \ref{F3}. The spheres' internal fields can be expanded as a multipole series with coefficients $\zeta^{(u)}_{l;0}$ which are related to to the scattering MECs $p_{l;0}^{(u)}$ in Eq.~\eqref{eq1} via $\zeta^{(u)}_{l;0} = p_{l;0}^{(u)}c_l^{(u)}/a_l^{(u)}$, where $a_l^{(u)}$ and $c_l^{(u)}$ are the Mie coefficients~\cite{Hoang2012}. The collective resonances are dominated by the internal  electric quadrupole (EQ) coefficients. 

Figure \ref{F3}(a) presents the magnitude and phase of the EQ coefficient representing the internal field of the $8$-th sphere for the two cases of $G =1.8$ nm and $G = 3.225$ nm. The spectral profiles of the magnitude, which agree with the $F_P$ evolution in Fig. \ref{F2}(c), show that the crossing of the two $L_{1,2}$ modes leads to the disappearance of the $L_2$ mode at $G = 3.225$ nm. In contrast to the magnitude plots, the phase plots always exhibit the signatures of all the three $L_{1,2,3}$ resonances. In the case of $G = 3.225$ nm, the disappearance of the $L_2$ resonance manifests itself as a strong phase fluctuation around the black $L_1$ peak, resulting in a small phase trough marked by the red $P_2$. This phenomenon is consistent with Feshbach's theory of two interfering resonances, where one resonance becomes sharper and the other becomes more lossy. Additionally, the phase jumps across the peaks in the spectral phase plots are not exactly $\pi$, due to coupling to other multipoles \cite{Friedrich1985}. At the enhancement and suppression points, the EQ coefficient becomes real, causing the phase changes of $2\pi$. 

Figures \ref{F3}(b) and \ref{F3}(c) present the physics of the super-cavity $L_1$ resonance from a far-field perspective. Without the resonance crossing ($G=0$ nm), the collective $L_1$ resonance exhibits high radiative losses, as shown in Fig. \ref{F3}(b), which displays $20$ circular fringes. Each fringes may be understood as a separate scattering loss channel. The far-field distribution of the super-cavity $L_1$ resonance in Fig.~\ref{F3}(c) shows a reduced number of only $8$ scattering loss channels, giving improved light trapping efficiency compared to the normal $L_1$ resonance. Remarkably, this non-vanishing far field of the super-cavity resonance is also a characteristic feature of Feshbach-type BICs--which include the Friedrich-Wintgen BICs--in the context of formal scattering theory \cite{Fonda1963,Hoang2022}.

\begin{figure}[tbp]
\includegraphics[width = \columnwidth]{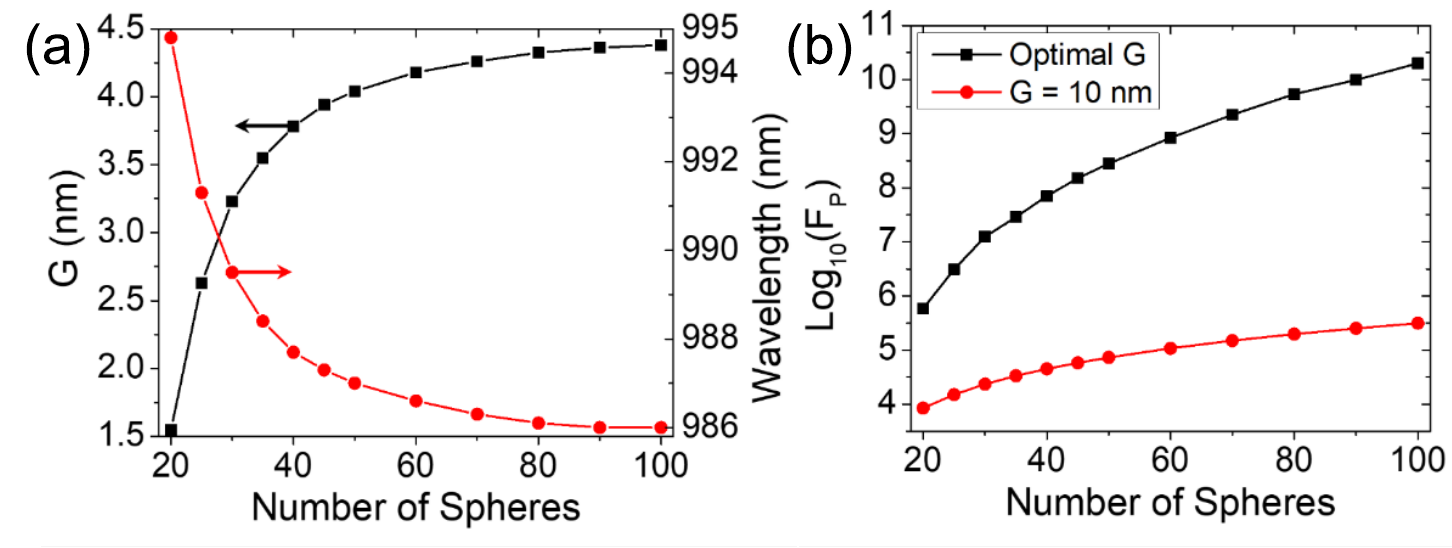}
\caption{\label{F4} Dependence of the super-cavity $L_1$ resonance on the number of spheres. (a) The optimal gap and its associated wavelength of the mode converge to $G = 4. 5$ nm and $\lambda = 986$ nm, respectively. (b) Purcell factor in logarithmic scale. Two plots associating with the super-cavity $L_1$ resonance (optimal $G$) and the normal collective $L_1$ resonance at $G = 10$ nm are presented for comparison.}
\end{figure}

{\it Flatband resonance.---} As the number of spheres is increased, the chain transitions into a quasi-one-dimensional (quasi-1D) photonic crystal (PC). Previous studies have shown that a bandwidth of this type of PC corresponds to the spectral range enclosing all the collective resonances of the same Mie modes~\cite{Hoang2020}. The single peak of the super-cavity $L_1$ resonance arising from the degeneracy of the collective Mie resonances suggests that its associated PC band is flat. To obtain a more accurate estimation of the bandwidths of the associated PC we consider chains of up to 100 spheres~\cite{Hoang2020}. 

\begin{figure}[htbp]
\includegraphics[width =0.75\columnwidth]{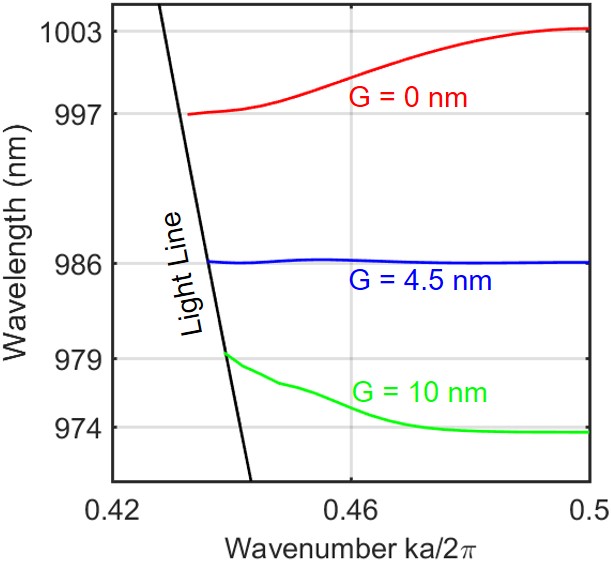}
\caption{\label{F5} Band structures for three different gap sizes indicate that the flatband occurs at $G = 4.5$ nm.}
\end{figure}

Figure \ref{F4}(a) shows the dependence of the super-cavity $L_1$ resonance on the number of spheres, demonstrating the convergence of its resonant wavelength and the corresponding $G$ towards $986$ nm and $4.5$ nm, respectively (see Supplementary Section S6 for further details). This suggests that a flatband of the PC with a period of 424.5 nm will occur at a single wavelength of $986$ nm, which is confirmed in Fig. \ref{F5}. In finite chains, the flatband is represented by the super-cavity $L_1$ resonance, which provides extreme enhancement of the LDOS as shown in Fig.~\ref{F4}(b). For comparison, the Purcell factor $F_P$ for the case of $G=10$ nm is also shown. The super-cavity $L_1$ resonances provide much larger enhancement compared to the normal $L_1$ resonances. For the chain of 100 spheres, the super-cavity resonance has a $Q$ factor of $6.6\times10^9$ and a giant $F_P$ factor of $>10^{10}$. Such high $F_P$ and $Q$ values are useful for the development of quantum networks \cite{Reiserer2015}, on-chip quantum light sources \cite{Liu2018}, and on-chip nanolasers \cite{Hoang2020}.      

Figure \ref{F5} displays the band structure of the quasi-1D PC for three different gap sizes obtained from a numerical solution of Maxwell's equations~\cite{Hoang2020}. 
All the three bands are situated below the light line, which characterizes them as guided PC modes. A remarkably flat band appears at $\lambda = 986$ nm for $G = 4.5$ nm, or equivalently $a = 424.5$ nm. The other two bands, with $G = 0$ nm and $G = 10$ nm, have bandwidths of 6 nm and 5 nm, respectively. The positions of the two bandedge $L_1$ modes for $G = 0$ nm and $G = 10$ nm, relative to their bands, reflect the crossing of the collective Mie resonances, as presented in Fig. \ref{F2}(c). This confirms that the origin of the flat band at $\lambda = 986$ nm is due to the degeneracy of the collective resonances existing below the light line. In the PC context, since the flatband is situated below the light line, the origin of the flatband is the degeneracy of a continuum of guided PC modes achieved via fine-tuning. Conventional wisdom expects the flatband to decouple from the far-field region. However, in the framework of MST, the flatband resonance does not vanish in the far-field region. This phenomenon is similar to photonic flatband resonances above the light line, as discussed in \cite{Yang2023}. Therefore, the multiple scattering theory provides a powerful framework for probing flatband physics from both PC and scattering perspectives. 



{\it Conclusion.---} We have elucidated the origin of the flatband resonances in the multiple multipole Mie scattering, demonstrating that they arise from a degeneracy of the collective resonances. Hybridizing the flatband resonance with an electric hotspot mode can lead to a giant enhancement of the Purcell factor of an electric dipolar emitter. Our findings not only shed new light on the physics of other flatband systems, including those based on moir\'e structures \cite{Wang2020,Mao2021}, but also suggest a smart optimization of nanophotonic quantum devices. Our multiple multipole scattering approach is applicable to cavities of arbitrary shapes, and it offers a powerful semi-analytical technique for investigating the quantum interaction between nanostructures and multipolar emitters \cite{Novotny2006}. Each multipole mode can also serve as a quasi-normal mode for the second quantization of the electromagnetic field in a nanophotonic system \cite{Sauvan2013,Kristensen2014,Franke2019,Medina2021}, providing a node in a quantum network \cite{Fabre2020}. 

\begin{acknowledgments}
This research is supported by A*STAR under its Career Development Fund (C210112012). D.L. acknowledges a support from the National Research Foundation, Singapore and A*STAR under its CQT Bridging Grant. Y.K. acknowledges a support from Australian Research Council (grant DP210101292) and the International Technology Center Indo-Pacific (ITC IPAC) via Army Research Office (FA520921P0034).
\end{acknowledgments}
\bibliography{Reference_PRL}

\end{document}


\setcounter{equation}{0}
\setcounter{figure}{0}
\setcounter{table}{0}
\setcounter{section}{0}

\makeatletter
\renewcommand{\theequation}{S\arabic{equation}}
\renewcommand{\thefigure}{S\arabic{figure}}
\renewcommand{\bibnumfmt}[1]{[S#1]}
\renewcommand{\citenumfont}[1]{S#1}
\renewcommand{\thesection}{S\arabic{section}}

\preprint{APS/123-QED}

\title{\textbf \Large Supplemental Material for \\
``Photonic Flatband Resonance in Multiple Light Scattering''}

\author{Thanh Xuan Hoang$^{1}$, Daniel Leykam$^{2}$, Yuri S. Kivshar$^{3}$}
\affiliation{$^{1}$Department of Electronics and Photonics, Institute of High Performance Computing A$^\star$STAR (Agency for Science, Technology and Research), 1 Fusionopolis Way, \# 16-16 Connexis, Singapore 138632, Singapore}%
\affiliation{$^{2}$Centre for Quantum Technologies, National University of Singapore, 3 Science Drive 2, Singapore 117543}%
\affiliation{$^{3}$Nonlinear Physics Center, The Australian National University, Canberra ACT 2601, Australia}%



\maketitle


\section{Multipole analyses of Mie and electric hotspot resonances} \label{S1}
\begin{figure}[htbp]
\includegraphics[width = 6.5 cm]{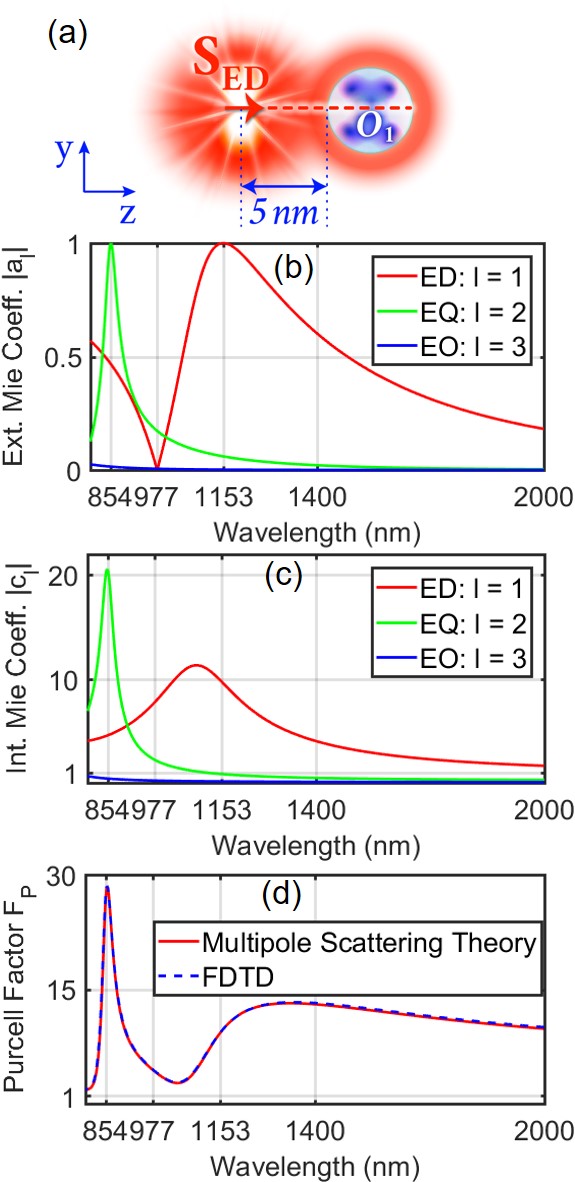}
\caption{\label{S1F1} (a) Interaction schematic of a $z$-oriented electric dipole and a sphere with a radius of $210$ nm and a refractive index of $3.5$. (b)-(c) Spectral profiles of the external and internal Mie coefficients associated with the sphere. (d) Purcell factor associates with the interaction between the dipole and the sphere.}
\end{figure}
Figure \ref{S1F1} presents the multipole analysis of the interaction between the $S_{ED}$ emitter and the single sphere. The spectral profiles of the ED, EQ, and EO Mie coefficients are shown in Figs. \ref{S1F1}(b) and \ref{S1F1}(c). Notably, we observe remarkable differences between the spectral profiles of the external and internal ED Mie coefficients. The asymmetrical lineshape of the external ED Mie coefficient arises from the interference of partial multipole waves resulting from multiple light scattering inside the sphere \cite{Hoang2012}. At $\lambda=1153$ nm, this interference is constructive, while at $\lambda =977$ nm, it is completely destructive. Although the constructive and destructive interference results in the asymmetrical lineshape, it is important to note that this is physically distinct from the Fano profile, which involves the interference between a resonance and a continuum \cite{Hoang2022}. On the contrary to the completely destructive interference of the external ED Mie coefficient at $\lambda =977$ nm, its internal counterpart is near optimal and hence originates from constructive interference of the partial multipole waves inside the sphere. For higher-order Mie coefficients, such as EQ and EO terms, the spectral profiles of the external and internal coefficients are similar in shape.

Figure \ref{S1F1}(d) shows the Purcell enhancement $F_P$ spectrum associated with the $S_{ED}$ emitter, which describes the enhancement of the emitter's emission rate due to its interaction with the nearby sphere. In our MST, we model that in the absence of the sphere, the emitter uniformly emits light into the vacuum across the spectral range. In the presence of the sphere, the emitter can emit light by two distinct mechanisms: (1) directly into the far-field region, and (2) via near-field coupling to the Mie resonances of the sphere. Interference between the Mie resonances and the direct emission continuum channel results in the asymmetrical spectral profile of $F_P$. This profile can be linked to the Fano profile in atomic physics \cite{Hoang2017}. Figure \ref{S1F1}(d) also presents the agreement between our calculation based on multipole scattering theory and that based on a numerical finite-difference time-domain (FDTD) simulation (Lumerical FDTD).               

\begin{figure}[htbp]
\includegraphics[width = 6.5 cm]{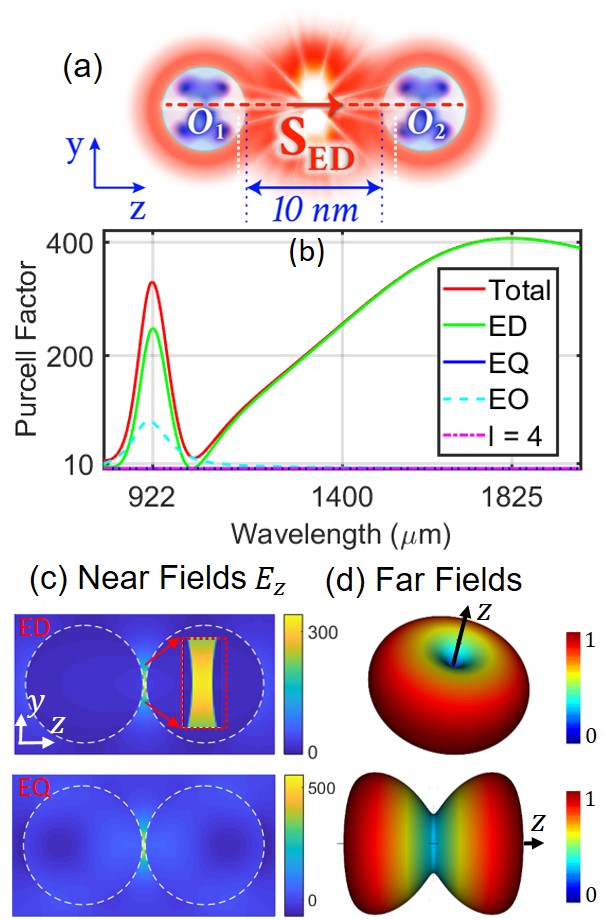}
\caption{\label{S1F2} (a) A schematic of the interaction between the $S_{ED}$ emitter and the two identical spheres. The spectral profile of the associated Purcell factor is presented in (b), which exhibits two resonant hotspot resonances. The near-field distribution and the far-field radiation pattern of the hotspot resonances are displayed in (c) and (d), respectively.}
\end{figure}
The schematic of the $S_{ED}$ emitter interacting with the two identical spheres is shown in Fig. \ref{S1F2}(a), along with the associated $F_P$ plot in Fig. \ref{S1F2}(b). The $F_P$ plot exhibits two resonances at $\lambda_{\text{EQ}} = 922$ nm and $\lambda_{\text{ED}} = 1825$ nm. We decompose the total $F_P$ into contributions due to partial multipole waves and find that the collective resonances of the two EDs behaves like a dipole wave in the far-field region, which is consistent with the radiation pattern shown in Fig. \ref{S1F2}(d). The contribution of the dipole wave is dominant in the case of the collective mode of the two EQs, although the resulting radiation pattern resembles that of an EQ mode. This highlights the need for caution when interpreting a collective resonance from far-field measurements. Figure \ref{S1F2}(c) shows the imaginary component of the near-field $E_z$ distribution for the two resonances. This indicates that the longitudinal field components are tightly confined in the gap. Due to the tight confinement of the electric hotspot resonance, the Purcell factor is enhanced by more than one order of magnitude compared to those shown in Fig. \ref{S1F1}(d).

\begin{figure}[htbp]
\includegraphics[width = 6.5 cm]{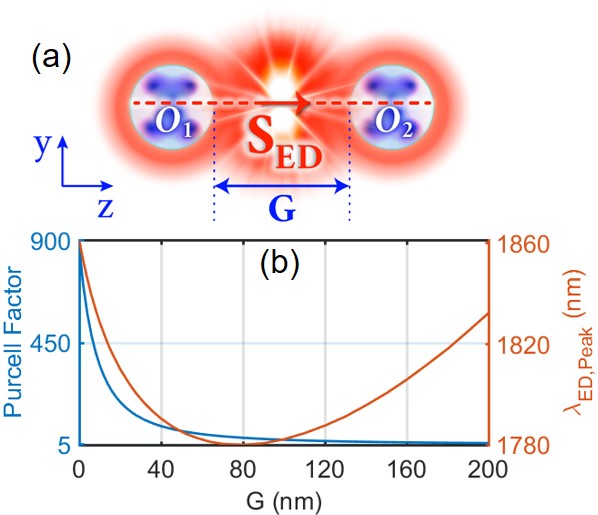}
\caption{\label{S1F3} (a) Schematic with the inter-sphere gap $G$. (b) Effect of the tuning gap $G$ on the resonant gap mode of the two EDs.}
\end{figure}
Figure \ref{S1F3} shows the effect of the gap size on the resonant ED electric hotspot mode. As the gap size decreases from 10 nm to 0 nm, the Purcell factor increases twofold. For gap sizes $G\leq 80$ nm, the peak wavelength decreases with increasing gap size, while the opposite trend is observed for $G > 80$ nm. In the latter case, the gap mode is weak enough that the two spheres can be treated as independent, and their interaction with the emitter is similar to that shown in Fig. \ref{S1F1}.

\section{Hybridization of electric hotspot and collective Mie resonances: $N = 3, 4, 5$} \label{S2} 
\begin{figure}[htbp]
\includegraphics[width = 6.5 cm]{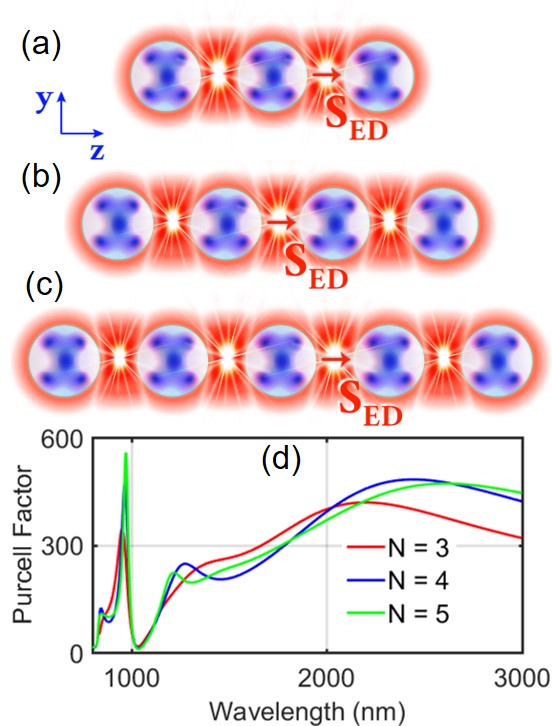}
\caption{\label{S2F1} (a)-(c) Schematics of the emitter interacting with three chains of spheres with sphere numbers of $N=3,4,5$, respectively. (d) The associated Purcell factor for these three cases.}
\end{figure}

Figures \ref{S2F1}(a)-(c) illustrate the three interaction schematics, while Fig. \ref{S2F1}(d) presents the effect of the sphere number $N$ on the corresponding $F_P$ spectrum. Increasing $N$ leads to a monotonically increasing peak associated with hybridization of the electric hotspot mode with the collective EQ resonance. However, this trend does not hold for the ED resonance; the electric hotspot mode does not hybridize with the collective ED resonance.    

\section{Multipole analyses of collective resonance and super-cavity modes: $N = 30$} \label{S3} 
\begin{figure}[htbp]
\includegraphics[width = 8 cm]{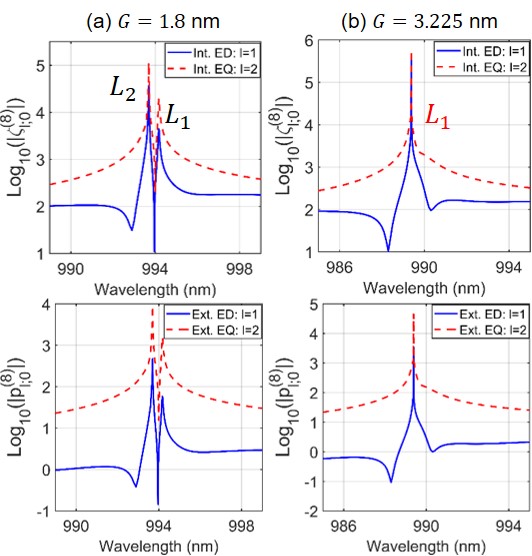}
\caption{\label{S3F1} Spectral profiles of the ED and EQ coefficients representing the internal (upper panel) and external (lower panel) fields associated with the $8$-th sphere for the two cases: (a) $G = 1.8$ nm and (b) $G = 3.225$ nm. Both the ED and EQ coefficients exhibit radiative loss.}
\end{figure}

Figure \ref{S3F1} presents the scattering $p_{l;0}^{(8)}$ and internal $\zeta_{l;0}^{(8)}$ ED and EQ expansion coefficients for the $8$-th sphere of the $30$-sphere chain with the two gap sizes: $G = 1.8$ nm and $G = 3.225$ nm. As shown in Fig. \ref{S3F1}(a), the super-cavity $L_2$ resonance is stronger than the normal $L_1$ resonance. This counterintuitive result is due to the crossing of the $L_3$ resonance, as discussed in the main text. Note that in our previous work \cite{Hoang2020}, the optimal gap sizes for the strongest collective resonances of the same Mie mode exist in the anti-crossing regime. In contrast, the strong enhancement of the $L_2$ mode in the present study occurs in the crossing regime. Figure \ref{S3F1}(b) shows that when $G$ is fine-tuned, $L_{1}$ and $L_2$ cross each other and degenerate into a single resonance. This super-cavity $L_1$ resonance also occurs in the crossing regime and is enhanced by over one order of magnitude compared to that in Fig. \ref{S3F1}(a).

It is worth noting that the crossing regime in which the super-cavity $L_{1;2}$ resonances occur is different from the anti-crossing regime of the matter Friedrich-Wintgen BICs \cite{Friedrich1985}. The latter has inspired the development of its photonic counterpart in the anti-crossing regime, also known as the avoided crossing regime \cite{Rybin2017}. However, the assumption made in the abstract coupled mode theory about the coupling between two closed channels and a single radiative channel \cite{Hsu2016} is inadequate for our current case, where the coupling between the $L_{1;2}$ resonances and the infinite number of radiation plane-wave channels supported by the surrounding 3D environment plays a crucial role. Therefore, the super-cavity resonance in the crossing regime observed in our study represents a new regime for photonic super-cavities and offers insights into the coupling mechanism between the collective resonances and the radiation channels in photonic systems. 

\begin{figure}[htbp]
\includegraphics[width = 8 cm]{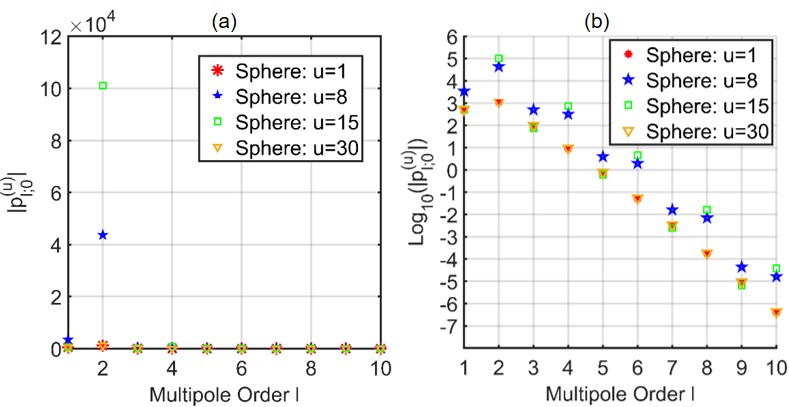}
\caption{\label{S3F2} Dominance of the EQ terms in the formation of the super-cavity $L_1$ resonance and the $10$ scattering MECs from different spheres. The MECs are displayed in linear (a) and log (b) scales.}
\end{figure}

Our multiple Mie scattering theory accounts for the multipole modes in both near- and far-field regions, allowing us to observe the consistency of spectral profiles in both the scattering and internal MECs. This consistency between near and far fields is crucial in the theory of resonant multiple light scattering \cite{Lagendijk1996}. It is important to distinguish our super-cavity resonances from those resulting from the strong interaction of two orthogonal modes of Fabry-Perot- and Mie-type, as presented in Ref. \cite{Rybin2017}. In that case, the two modes do not interact in the near-field region inside the single particle, but interact strongly in the far-field region to create the super-cavity resonance. Our super-cavity resonance, on the other hand, occurs in the crossing regime due to both near- and long-range couplings between $600$ multipole terms that describe the near and far fields of the $30$ spheres. Each sphere supports $10$ scattering and $10$ internal multipole terms, and the ten scattering MECs for different spheres at the peak wavelength in the case of $G = 3.225$ nm are shown in Fig. \ref{S3F2}.

Figure \ref{S3F2}(a) shows that the scattering EQ ($l=2$) terms dominate the scattering fields from the four spheres, most notably the $15$th middle sphere. This is in contrast to the physics of Fabry-Perot cavity resonances \cite{Hoang2020}, where radiation loss occurs only at the ends of the chain. Our study reveals that, for the super-cavity $L1$ resonance, the radiation loss is actually dominated by scattering from the middle spheres ($u=15$). These findings have implications for the validity of the tight-binding model in studying the physics of quasi-BIC modes \cite{Sidorenko2021}. 

Figure \ref{S3F2}(b) shows that the scattering MECs, except for the scattering EQ coefficients, decrease exponentially with increasing multipole order $l$. However, all of the multipole orders are non-zero and contribute to the physics of the super-cavity resonances. In fact, the truncated multipole order of $L_u$ appearing in Eq. (1) has been a topic of substantial interest \cite{Hoang2017,Hoang2022}. We showed in Ref. \cite{Hoang2022} that even for studying the collective resonances of the lowest order ED and MD resonances, we need to include $7$ MECs to obtain results that match numerical simulations using the FDTD technique \cite{Hoang2022}. This is in contrast to some models of multiple light scattering, which use only one \cite{Kornovan2021} or two \cite{Huang2011} multipole eigenfunctions to describe the scattering fields. These findings suggest that the inclusion of partial multipole fields with higher orders may be necessary for designing and optimizing photonic Mie-type structures.

\begin{figure}[htbp]
\includegraphics[width = 8 cm]{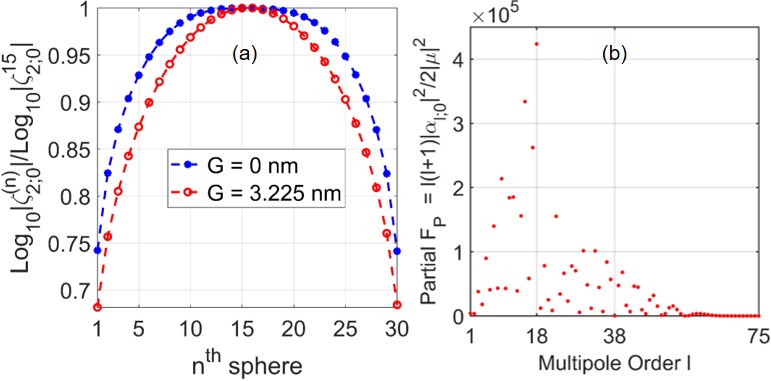}
\caption{\label{S3F3} (a) Internal EQ coefficients decay from that of the middle sphere more rapidly for the super-cavity $L_1$ resonance than for the normal collective $L_1$ resonance, indicating stronger localization. (b) Partial $F_P$ decomposition according to Eq. (4) for the super-cavity $L_1$ resonance, providing insight into the relative contributions of different multipole terms to the overall field distribution.}
\end{figure}
Figure \ref{S3F3}(a) shows the decay of the internal EQ coefficients, which correspond to the decay of the internal fields of the 30 spheres. We observe that the super-cavity $L_1$ resonance with $G = 3.225$ nm decays faster than the normal $L_1$ resonance with $G=0$ nm, but the decay is not exponential. An exponential decay would be represented by a straight line in Fig. \ref{S3F3}(a). Figure \ref{S3F3}(b) displays the partial contributions from all $75$ partial multipole fields in Eq. (3), where $L_{\text{tot}}=75$. We formulate the partial contributions to the total $F_P$ as shown in Eq. (4). The radiation power from the whole system consisting the ED emitter and the 30-sphere chain approaches the far-field region as many partial multipole fields of different MECs. Fine-tuning $G$ will redistribute these MECs to result in the change of the far-field pattern, as discussed in the main text.   

\section{Critical roles of long-range and off-diagonal couplings: $N = 30$}\label{S4}
\begin{figure}[htbp]
\includegraphics[width = 7 cm]{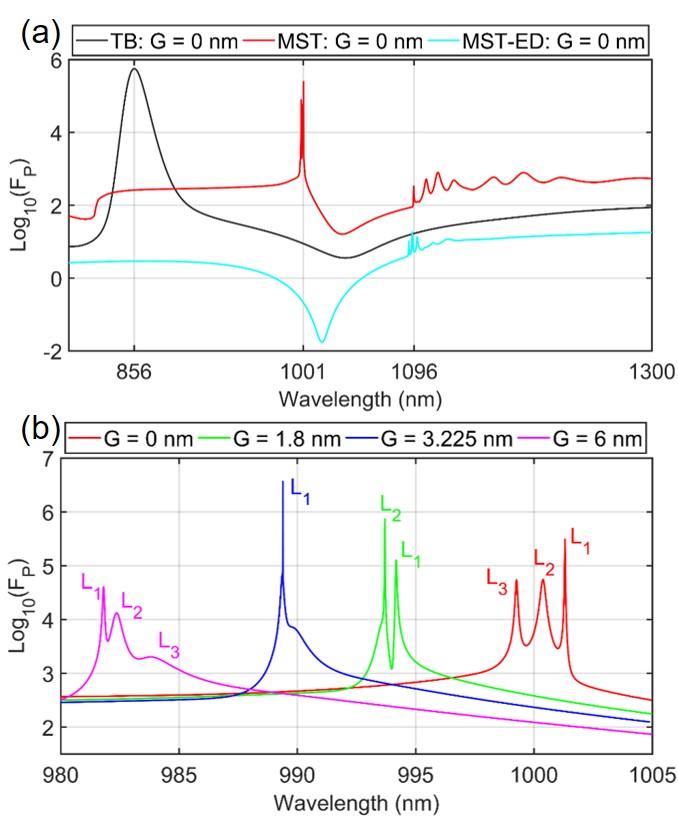}
\caption{\label{S4F1} (a) The spectral $F_P$ plot from three different models: (1) tight binding (TB), (2) exact multiple scattering theory (MST), and (3) MST-ED. The TB and MST-ED models fail to capture the physics of the collective resonances as exhibited by the exact MST model. (b) The impact of fine-tuning the gap size $G$ on the crossing of the collective resonances.}
\end{figure}

Figure \ref{S4F1} provides additional insights into the physics of collective resonances shown in Fig. 2 of the main text. Figure \ref{S4F1}(a) shows the inadequacy of both the TB and MST-ED models in accurately describing both the collective ED and EQ resonances. While the MST-ED model predicts the existence of collective ED resonances, it underestimates the magnitude of the LDOS enhancement due to the neglect of the off-diagonal coupling between the ED modes and other higher-order multipole terms. Moreover, neither the TB nor the MST-ED models can predict the crossing of the collective resonances as observed in the exact MST model. Figure \ref{S4F1}(b) presents the crossing regime of the collective resonances based on the exact MST model. Therefore, both the long-range and off-diagonal couplings are essential to the underlying physics of collective resonances and flatband modes.

\section{Crossing regime of the collective resonances: $N = 30$}\label{S5}
\begin{figure}[htbp]
\includegraphics[width = 7.5 cm]{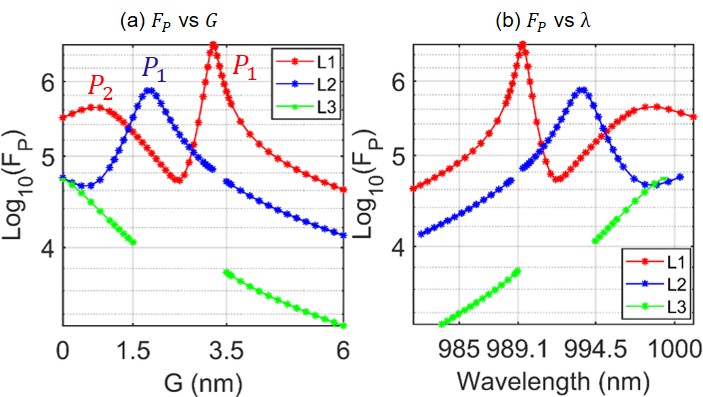}
\caption{\label{S5F1} By fine-tuning the gap size $G$, we can cause the three collective modes supported by the 30-sphere chain to cross. In (a), we show that the $L_3$ mode disappears for $G\in (1.5,3.5)$ nm. In (b), we plot the dependence of $F_P$ and the corresponding wavelength on the fine-tuning gap $G$.}
\end{figure}

Figure \ref{S5F1} presents the same data as Fig. 2 in the main text, but in a different format. In the figure, the blue curve exhibits a single peak, marked by the blue $P_1$, which we refer to as the super-cavity $L_2$ resonance. Within the fine-tuning range $G\in (0,1.8)$ nm, the associated $F_P$ of this resonance increases by over one order of magnitude. The red curve in the figure has two peaks, marked by the red $P_1$ and $P_2$, where only the $P_1$ peak arises from the crossing of the $L_1$ and $L_2$ resonances. The physics of the $P_2$ peak is similar to that in our previous work on optimizing the $L_1$ resonance, where the optimal resonance occurs in the anti-crossing regime \cite{Hoang2020}.

At the highest peak of the super-cavity $L_1$ resonance, both the $L_2$ and $L_3$ resonances vanish. If we extend the chain to infinity and fine-tune the gap $G$ such that all the collective resonances cross the band-edge $L_1$ resonance, we obtain a super-cavity $L_1$ resonance corresponding to a flat band of the quasi-1D PC. In other words, the flatband arises from the degeneracy of collective Mie resonances into the PC band-edge $L_1$ mode.

\section{Divergence and convergence of the flatband with $N$}\label{S6}
\begin{figure}[htbp]
\includegraphics[width = 8 cm]{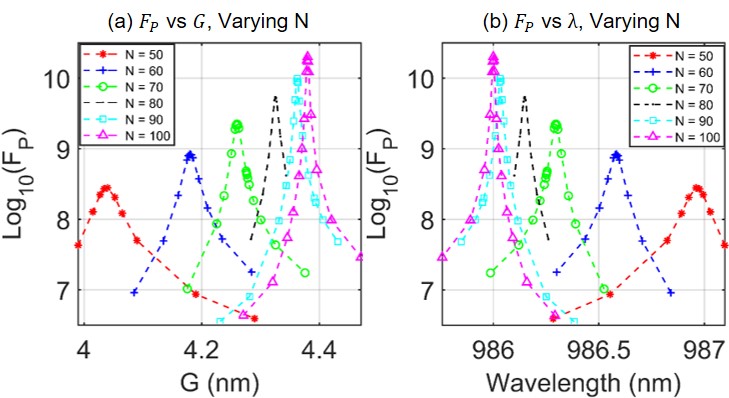}
\caption{\label{S6F1} Divergence and convergence of the flatband (super-cavity $L_1$) resonance with an increasing number of spheres. (a) The divergence of the $F_P$ peak and the associated convergence of the optimal gap $G$ with the increasing number of spheres. (b) The wavelength of the flatband converges to $\lambda = 986$ nm when $N\rightarrow 100$.}
\end{figure}

Figures \ref{S6F1}(a)-(b) show the optimal gap and corresponding wavelength of the flatband resonance for the six chains with different $N$. While the $F_P$ peak value diverges with the increasing number of spheres, the optimal gap and its corresponding wavelength converge to $G = 4.5$ nm and $\lambda = 986$ nm, respectively.   
\bibliography{SupMat}